# Giant magnetocaloric effect in magnetoelectric $Eu_{1-x}Ba_xTiO_3$

Km Rubi, Pawan Kumar, D. V. Maheswar Repaka, Ruofan Chen, Jian-Sheng Wang, and R. Mahendiran[*]

Department of Physics, 2 Science Drive 3, Faculty of Science, National University of Singapore, Singapore 117551, Republic of Singapore

## Abstract

We report the magnetic entropy change ($\Delta S_m$) in magnetoelectric $Eu_{1-x}Ba_xTiO_3$ for $0.1 \leq x \leq 0.9$. We find $-\Delta S_m = 11$ (40) J/kg·K in $x = 0.1$ for a field change of 1 (5) Tesla respectively, which is the largest value among all Eu-based oxides. $\Delta S_m$ arises from the field-induced suppression of the spin entropy of $Eu^{2+}$:$4f^7$ localized moments. While $|-\Delta S_m|$ decreases with increasing $x$, $|-\Delta S_m| = 6.58$ J/kg·K observed in the high spin diluted composition $x = 0.9$ is larger than that in many manganites. Our results indicate that these magnetoelectrics are potential candidates for cryogenic magnetic refrigeration.

---

[*] corresponding author (phyrm@nus.edu.sg)

**I. Introduction**

There has been intense research to discover new magnetic materials displaying large magnetocaloric effect (MCE) for magnetic cooling applications.[1] The MCE refers to cooling of a paramagnetic or ferromagnetic (antiferromagnetic) substance upon adiabatic demagnetization (magnetization), and the process is accompanied by a magnetic entropy change. Magnetic refrigeration is a reliable and commercially available technique to reach millikelvin temperature that exploits magnetic entropy change of non-interacting spins of paramagnetic salts. For near room-temperature magnetic refrigeration, ferromagnets are preferred since the entropy change $\Delta S_m$ attains a maximum value at the phase transition temperature.[2] Materials such as doped manganites, e.g., $La_{0.7-x}Pr_xCa_{0.3}MnO_3$,[3] $La(Fe,Si,Mn)H_\delta$,[4] Mn-Fe-P-Si,[5] and Ni-Mn-In alloys[6] are actively being considered for applications in the temperature range of 200 - 350 K. Since a single material for magnetic refrigeration is restricted in temperature range, tailored magnetic materials within a family are sought after because related properties such as thermal conductivity, heat capacity, and electrical conductivity may show a systematic variation with changing compositions. Despite the current race for near room-temperature refrigeration, there has been a need to find new and efficient magnetic refrigerants suitable for cryogenic magnetic cooling applications (e.g. liquefaction of hydrogen, infrared bolometer) in the temperature range of 1 - 20 K.[7] Superparamagnetic molecular magnetic clusters and gadolinium garnet ($Gd_3Ga_5O_{12}$) are considered as potential magnetic refrigerants in this temperature range.[8]

In this work, we report the occurrence of a giant magnetocaloric effect at cryogenic temperatures ($T < 30$ K) in a novel class of magnetoelectric materials, $Eu_{1-x}Ba_xTiO_3$, whose end members have distinct ferroic orders. While $EuTiO_3$ ($x = 0$) is a paraelectric-antiferromagnet, $BaTiO_3$ ($x = 1$) is a ferroelectric-nonmagnet.[9] $EuTiO_3$ orders antiferromagnetically in G-type structure below $T_N$ = 5.5 K due to long-range order of Eu$^{2+}$:4f$^7$ (J = S = 7/2) spins, while the ferroelectric active Ti$^{4+}$(d$^0$) ions remain non-magnetic. The absence of ferroelectricity in bulk EuTiO$_3$ is attributed to the influence of quantum fluctuations on a specific phonon mode responsible for ferroelectricity.[10] The antiferromagnetism in EuTiO$_3$ is also unusual because the antiferromagnetic superexchange interaction mediated via Eu-4f- Ti-3d hybridized state in this compound seems to dominate over the usual superexchange interaction mediated via oxygen ion in perovskite oxides.[10] EuTiO$_3$ is a topic of immense interests in recent years due to the discoveries of magnetocapacitance effect in single crystal,[11] tensile-stress induced ferromagnetism and ferroelectricity in thin film[12], and electric field modulation of tetragonal domain orientation.[13]

Nearly four decades ago, D. L. Janes et al.[14] suggested that the solid solution Eu$_{0.5}$Ba$_{0.5}$TiO$_3$ would be simultaneously ferroelectric and ferromagnetic. In recent experiments, K. Z. Rushchanskii et al.[15] found ferroelectric hysteresis loop around 130 K and antiferromagnetism below $T_N$ = 1.9 K in the same composition. While H. Wu et al.[16] investigated theortically the intrinsic coupling between magnetism and dielectric properties in Eu$_{1-x}$Ba$_x$TiO$_3$ (0 ≤ $x$ ≤ 0.2), T. Wei et al.[17] reported dielectric constant and electrical polarization in polycrystalline samples (0 ≤ $x$ ≤ 1). It was found that while $x$ ≤ 0.15 is paraelectric, ferroelectricity develops in $x$ > 0.2 as evidenced by ferroelectric hysteresis loop. Since Eu$^{2+}$ has a large spin magnetic moment of µ = 7µ$_B$ with a zero orbital angular momentum (L = 0), these compounds may show a large isotropic magnetic entropy change.[18] However, systematic studies of magnetization and magnetocaloric effect in Ba doped samples of EuTiO$_3$ have not been reported so far. Eu$_{1-x}$Ba$_x$TiO$_3$ also provides a unique opportunity to study how magnetocaloric effect changes with the dilution of the rare earth site. These considerations motivated the present work.

## II. Experimental Details

Polycrystalline $Eu_{1-x}Ba_xTiO_3$ ($0 \leq x \leq 0.9$) samples were synthesized through solid state reaction method. The powders of $Eu_2O_3$, $BaCO_3$, and $TiO_2$ were mixed in the stoichiometric ratio. After mixing and grinding, powders were sintered at 1200°C for 24 hours in 95% Ar and 5% $H_2$ atmosphere (which reduces $Eu^{3+}$ to $Eu^{2+}$). After two intermediate grinding and heating at 1200°C, powders were pressed in a uniaxial press into a pellet and the pellet was sintered at 1300°C for 24 hours in the same atmosphere. X-ray diffraction (XRD) done at room temperature using Cu K$\alpha$ radiation confirmed that samples are single phase. Magnetization was measured using a Physical Property Measurement System (PPMS) equipped with vibrating sample magnetometer (VSM) probe.

## III. Result and Discussion

We show powder X-ray diffraction pattern of $Eu_{1-x}Ba_xTiO_3$ ($0 \leq x \leq 0.9$) in the main panel of Fig. 1. Compositions $x = 0$ to 0.7 are cubic but $x = 0.8$ and 0.9 show tetragonal symmetry as evidenced by the splitting of (200) diffraction peak into (002) and (200) peaks in the latter two compounds, in agreement with the results of T. Wei *et al*. Rietveld analysis on X-ray data was performed to obtain lattice parameters. Inset shows *a* and *c* lattice parameters as a function of composition. The *a* parameter increases linearly from 3.904 Å for $x = 0$ to 3.975 Å for $x = 0.9$ and the c parameter is 3.977 Å and 4.021 Å for $x = 0.8$ and 0.9, respectively. Increase of lattice constant *a* with increasing *x* is due to bigger ionic radius of $Ba^{2+}$ compared to $Eu^{2+}$ ions.

The main panel of Fig. 2(a) shows the temperature dependences of magnetization $M(T)$ for $x = 0.1 - 0.9$ measured upon cooling from 300 K to 2.5 K under a magnetic field of $H = 1$ kOe and the inset compares $M(T)$ of $x = 0.1$ and 0.2. We have shown the data only in the low temperature range ($T$ = 2.5 K to 25 K) for clarity. The prominent peak around $T = T_N$ = 3.47 K in $x$ = 0.1 (see the inset) indicates the onset of antiferromagnetic ordering, which is lower than that of EuTiO$_3$ ($T_N$ = 5.5 K).[10] $T_N$ decreases to 2.79 K and the magnitude of $M$ below 5 K increases for $x$ = 0.2. As $x$ increases above 0.2, $T_N$ either shifts below our measurement limit of $T$ = 2.5 K or 4f spins become disordered. The reduction of $T_N$ is much faster than a linear behavior of 5.5(1-$x$) K, which is expected to be valid for small $x$. The main panel shows that the value of $M$ at the lowest temperature increases up to $x$ = 0.3 and then starts decreasing under $H$ = 0.1 T. Fig.2 (b) shows the field dependence of magnetization, $M(H)$ measured at 2.5 K from 0 to 5 T. None of the sample shows hysteresis. $M$ increases linearly with $H$ up to 1 T for $x$ = 0.1 as the spin configuration changes from antiferromagnetic to spin flop state and angle between the flopped spins decreases towards zero as the field increases further leading to an induced ferromagnetic state. In the field range 0 T < $H$ < 1 T, $M(H)$ of $x$ = 0.1 crosses over the $M(H)$ curve of $x$ = 0.2 because of its higher $T_N$. $M(H)$ curves for $x$ > 0.2 resemble that of a soft ferromagnet. However, absence of the hysteresis and remanence suggests that these samples are most likely in the paramagnetic state in zero field, but ferromagnetic order is induced by the external magnetic field aided by low thermal energy. Inset of Fig. 2(b) compares the experimental data at 5 T with the theoretically expected saturation values according to $(7/2)(1-x)g\mu_B$/f.u, assuming a Landé factor of $g$ = 2. The saturation magnetization at 5 T, $M_{sat}$, decreases gradually with increasing $x$ from 6.05 $\mu_B$/f.u. to 0.7 $\mu_B$/f.u. for $x$ = 0.1 to 0.9. The spin only theoretical moments closely match with the experimental results.

Fig. 3(a) shows the inverse susceptibility ($\chi^{-1}$) of all the compositions, including $x$ = 0. It can be seen that $\chi^{-1}$ shows a change of slope at a specific temperature $T^*$ (marked by the arrow) for each composition though it is more pronounced at higher Ba contents ($x$ = 0.8 and 0.9). For clarity, we show the data for $x$ = 0, 0.3, and 0.4 in the inset. No systematic dependence of $T^*$ with $x$ is found. It is likely that this anomaly is caused by structural transition driven by antiferro-distortive rotation of $TiO_6$ octahedra, as seen in the high resolution synchrotron diffraction experiment on $EuTiO_3$.[13] We show the low temperature ($T$ = 2.5 – 35 K) behavior of the inverse susceptibility in Fig. 3(b) and it is fitted with the Curie-Weiss law $\chi^{-1} = (T + \theta_p)/C$ where $\theta_p$ is the paramagnetic Curie temperature and $C$ is the Curie constant, which is related to the effective magnetic moment ($\mu_{eff}$) of $Eu^{2+}$ ions in the paramagnetic state. We show $\theta_p$ and $\mu_{eff}$ in the inset of Fig. 3(b). It is found that $\theta_p$ is positive for all the compositions and its magnitude decreases with increasing $x$ ($\theta_p$ = 3.05 K for $x$ = 0.1 to $\theta_p$ = 0.085 K for $x$ = 0.9). The effective magnetic moment decreases from 7.37 $\mu_B$ for $x$ = 0.1 to 3.56 $\mu_B$ for $x$ = 0.9 due to $Eu^{2+}$ site dilution by $Ba^{2+}$ ions.

Fig. 4(a)-(d) show $M(H)$ isotherms obtained at different temperatures for four selected compositions ($x$ = 0.1, 0.3, 0.5, and 0.9). Although $T_N$ of $x$ = 0.1 is 3.47 K, $M$ increases nonlinearly with $H$ up to ~ 24 K in the paramagnetic state and linear $M$-$H$ dependence is seen only above 30 K. The nonlinear behavior of $M(H)$ in the paramagnetic state could arises from the fact that the ratio of the Zeeman energy to thermal energy is $gS\mu_B H/k_B T$=2.35 at $H$ = 5 T and $T$ = 10 K for S = 7/2 and hence the magnetic field induces ferromagnetic ordering $Eu^{2+}$ spins even above $T_N$. Similar behavior is seen in other compositions (e.g., $x$ = 0.2) as well. For $x$ = 0.9, this ratio is only 0.2 and hence it behaves like a paramagnet. Classical (Langevin) or quantum mechanical model of paramagnetism predicts magnetization curves measured at different temperatures should fall on a single curve when $M$ is plotted against $H/T$. We show $M$ versus $H/T$ for $x$ = 0.1-0.9 in Fig. 4(e)-(h). For the highly spin diluted composition $x$ = 0.9, all the curves almost fall on a single master curve. We have fitted the experimental data using the mean-field expression for the magnetization given by von Ronke et al.[18] The model fits the experimental data perfectly if we assume the parameters 0.21$J_1$, 0.21$J_2$ and a scaling factor of 1.02 for the magnetization, where $J_1/k_B$ = -0.037 K and $J_2/k_B$ =

0.069 K are the values of the nearest neighbor and the next nearest neighbor interactions, respectively. Detail analysis is beyond the scope of this paper and will be presented elsewhere. With decreasing $x$, deviation from high temperature curves occurs below 30 K. It can be seen that for a given $H/T$ value, magnitude of $M$ increases with lowering temperature and it is larger for smaller the $x$. This can be attributed to the increasing interaction between 4f spins with lowering temperature or with increasing magnetic field. We have also plotted $M^2$ versus $H/M$ isotherms (known as the Arrot plots) in the inset. The positive slope of the Arrot plot suggests that the paramagnetic to antiferromagnetic phase transition in $x = 0.1$ is second order. For a normal second order paramagnetic to ferromagnetic transition, linear extrapolation of higher field $M^2$ versus $H/M$ line is expected to intercept the origin at $T = T_C$. However, we do not find such a trend in our samples because the phase transition has not taken place within the measured temperature range.

From the measured magnetization isotherms, we can calculate the magnetic entropy change $\Delta S_m = S_m(H)-S_m(0)$ using numerical integration of the Maxwell's thermodynamic relation $\Delta S_m = \int_0^H \left(\frac{\partial M}{\partial T}\right)_H dH$. Field induced ordering of $4f^7$ spins of $Eu^{2+}$ ions alone is responsible for the magnetic entropy change in the studied compounds. We plot the temperature dependence of -$\Delta S_m$ for six samples ($x = 0.1, 0.2, 0.3, 0.5, 0.7$ and $0.9$) in Fig. 5(a)-(f). When $\Delta H = 0.5$ T, -$\Delta S_m$ of $x = 0.1$ is nearly zero above 50 K, but it increases with lowering temperature and shows a peak at $T = 4.5$ K where it reaches a maximum value of 4.3 J/kg·K. The peak value of -$\Delta S_m$ increases with the increasing value of $\Delta H$ (-$\Delta S_m$ = 11.60, 21.89, 31.46, and 36.12 J/kg·K for $\Delta H$ = 1, 2, 3, and 4 T, respectively) and finally it reaches 40 J/kg·K for $\Delta H = 5$ T. The position of the peak shows negligible shift (< 0.03 K) as the field changes from 0.5 T to 5 T. The observed value of the magnetic entropy change is higher than the maximum value of -$\Delta S_m$ = 16 J/kg·K for $\Delta H = 5$ T found for R = Dy among the rare earth titanates RTiO$_3$ (R = Dy, Ho, Er, Tm Yb) series.[19] The observed -$\Delta S_m$ value is also higher than the maximum values reported in other Eu based materials such as EuO (17.5 J/kg·K)[20], Eu$_3$O$_4$ (12.7 J/kg·K)[21], EuDy$_2$O$_4$ (23 J/kg·K)[22], Eu$_8$Ga$_{16}$Ge$_{30}$-EuO composite (11.2 J/kg·K)[23], Eu$_{0.45}$Sr$_{0.55}$MnO$_3$ (7 J/kg·K) showing a first-order transition[24] but comparable to EuSe (37.5 J/kg·K)[25] and EuS (~38 J/kg·K)[26] for the same field change. The peak also occurs in $x$ = 0.2 at $T$ = 3.5 K but other compositions do not exhibit a peak since $T_N$ decreases below the minimum temperature of 2.5 K reachable in our cryostat. The maximum value of -$\Delta S_m$ at the lowest temperature decreases with increasing Ba content. However, even in the most diluted sample ($x$ = 0.9), -$\Delta S_m$ reaches 6.58 J/kg·K for $\Delta H$ = 5 T, which is higher than -$\Delta S_m$ = 1 - 4 J/kg·K for the same field strength found in the majority of manganites exhibiting second order paramagnetic to ferromagnetic transitions.[27]

Fig. 6(a) shows the field dependence of -$\Delta S_m$ at $T$ = 5.5 K for all the compositions. As we can see, -$\Delta S_m$ increases superlinearly with increasing magnetic field for all the compositions and the largest change occurs for the $x$ = 0.1 sample. We plot -$\Delta S_m$ as a function of Ba content ($x$) at five different temperatures in Fig. 6(b). -$\Delta S_m$ at $T$ = 5.5, 9.5, 15 and 26 K decreases nearly linearly with increasing $x$ whereas -$\Delta S_m$ at $T$ = 2.75 K decreases below $x$ = 0.3 due to the presence of antiferromagnetism in these samples.

## IV. Summary

We have found a giant magnetic entropy change varying from -$\Delta S_m$ = 40 J/kg·K to 6.7 J/kg·K at $T$ = 4.5K for $\Delta H$ = 5 T in multiferroic Eu$_{1-x}$Ba$_x$TiO$_3$ (0.1 ≤ $x$ ≤ 0.9) compounds. $\Delta S_m$ arises from the field induced suppression of the spin entropy of Eu$^{2+}$:4f$^7$ moments and the observed values are larger than other Eu-based oxides. The absence of hysteresis in the field dependences of magnetization and the magnetic entropy change is an added advantage of this series of compounds. In view of the observed giant magnetocaloric effect, these compounds may be of interest for cryogenic magnetic refrigeration below 20 K. However, direct measurement of the adiabatic temperature is highly desirable for practical consideration. Because magnetism and ferroelectricity coexist in certain compositions, it will be fascinating to investigate the possibility of magnetic tunable electrocaloric effect in these materials.

**Acknowledgment:** R. M. thanks the Ministry of Education, Singapore for supporting this work (Grant no. R144-000-308-112).

**Figure Caption:**

**Fig 1**. Powder X-ray diffraction pattern of $Eu_{1-x}Ba_xTiO_3$ ($0 \leq x \leq 0.9$) at room temperature. While $x \leq 0.7$ are cubic, $x = 0.8$ and $0.9$ are tetragonal. Inset shows the lattice parameters as a function of Ba content ($x$).

**Fig 2.** (a) Temperature dependence of magnetization ($M$) of $Eu_{1-x}Ba_xTiO_3$ samples for $x = 0.1$-$0.9$. Inset shows $M(T)$ for $x = 0.1$ and $0.2$. $T_N$ is the Néel temperature. (b) Field dependence of $M$ at 2.5 K for all compositions ($x$). Inset shows experimental value the saturation magnetization ($M_{sat}$) at 5 T (closed square) and calculated value (open circle) assuming $gS = 7$.

**Fig 3.** (a) Temperature dependence of the inverse susceptibility ($\chi^{-1}$) for different compositions ($x$). Arrows indicate the occurrence of possible structural transitions. Inset: $\chi^{-1}$ ($T$) for $x = 0.1$, $0.3$ and $0.4$. (b) $\chi^{-1}$ ($T$) and the Curie-Weiss fit in the low temperature range for all the compositions. Inset: Composition dependence of the paramagnetic Curie temperature ($\theta_p$) and the effective magnetic moment ($\mu_{eff}$) obtained from the Curie constant.

**Fig 4**. Left column: Magnetization isotherms at different temperatures for (a) $x = 0.1$, (b) 0.3, (c) 0.5, and (d) 0.9. Right column: M versus $\mu_0 H/T$ graphs for (e) $x = 0.1$, (f) 0.3, (g) 0.5, and (h) 0.9. Arrot plots ($M^2$ versus $H/M$ graphs) are shown as insets. Solid lines in the main panel of Fig.4(h) indicate the fits obtained using the mean field model.

**Fig 5**. Temperature dependence of the magnetic entropy change ($-\Delta S_m$) for (a) $x = 0.1$, (b) 0.2, (c) 0.3, (d) 0.5, (e) 0.7 and (f) 0.9 for a field change of $\Delta H = 0.5, 1, 2, 3, 4$ and $5$ Tesla.

**Fig 6.** (a) Field dependence of $-\Delta S_m$ at 5.5 K for all the compositions ($x$). (b) Composition ($x$) dependence of $-\Delta S_m$ at $T = 2.75$ K, $5.5$ K, $9.5$ K, $15$ K and $26$ K.

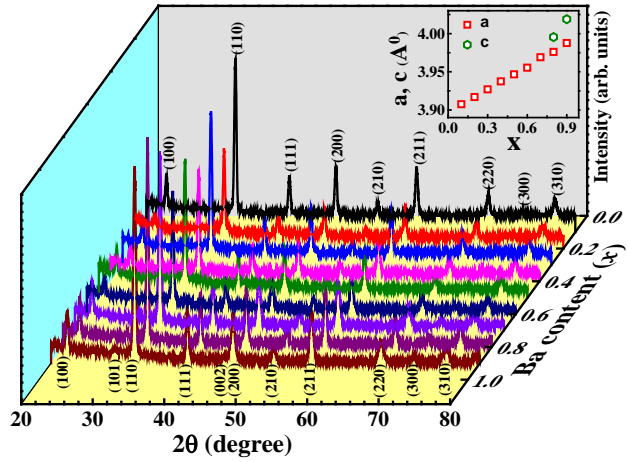

Fig.1 Km Rubi *et al.*

**Fig 1**. Powder X-ray diffraction pattern of $Eu_{1-x}Ba_xTiO_3$ ($0 \leq x \leq 0.9$) at room temperature. While $x \leq 0.7$ are cubic, $x = 0.8$ and $0.9$ are tetragonal. Inset shows the lattice parameters as a function of Ba content ($x$).

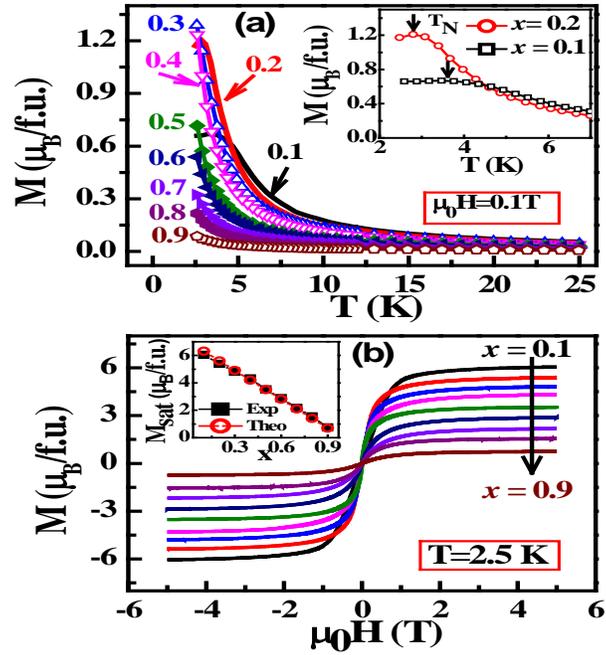

Fig 2. Km Rubi et al.

**Fig 2.** (a) Temperature dependence of magnetization (*M*) of $Eu_{1-x}Ba_xTiO_3$ samples for $x$ = 0.1-0.9. Inset shows *M(T)* for $x$ = 0.1 and 0.2. $T_N$ is the Néel temperature. (b) Field dependence of *M* at 2.5 K for all compositions (*x*). Inset shows experimental value the saturation magnetization ($M_{sat}$) at 5 T (closed square) and calculated value (open circle) assuming gS = 7.

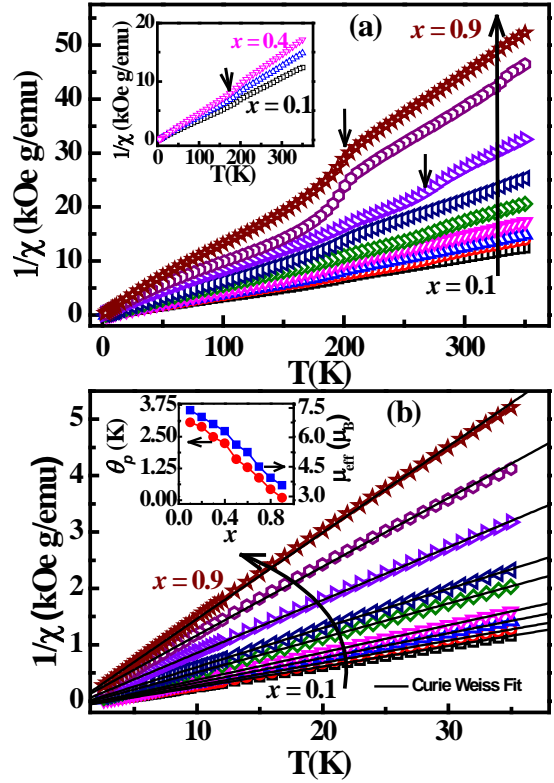

Fig. 3 Km Rubi *et al*.

**Fig 3.** (a) Temperature dependence of the inverse susceptibility ($\chi^{-1}$) for different compositions (*x*). Arrows indicate the occurrence of possible structural transitions. Inset: $\chi^{-1}$ (*T*) for *x* = 0.1, 0.3 and 0.4. (b) $\chi^{-1}$ (*T*) and the Curie-Weiss fit in the low temperature range for all the compositions. Inset: Composition dependence of the paramagnetic Curie temperature ($\theta_p$) and the effective magnetic moment ($\mu_{eff}$) obtained from the Curie constant.

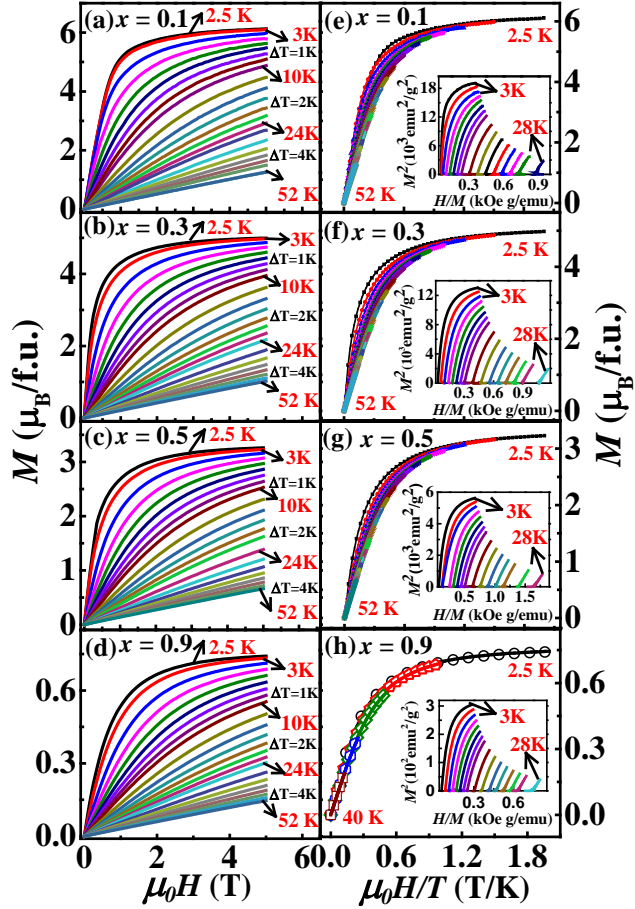

Fig.4 Km Rubi *et.al.*

**Fig 4**. Left column: Magnetization isotherms at different temperatures for (a) $x = 0.1$, (b) 0.3, (c) 0.5, and (d) 0.9. Right column: M versus $\mu_0H/T$ graphs for (e) $x = 0.1$, (f) 0.3, (g) 0.5, and (h) 0.9. Arrot plots ($M^2$ versus $H/M$ graphs) are shown as insets. Solid lines in the main panel of Fig.4(h) indicate the fits obtained using the mean field model.

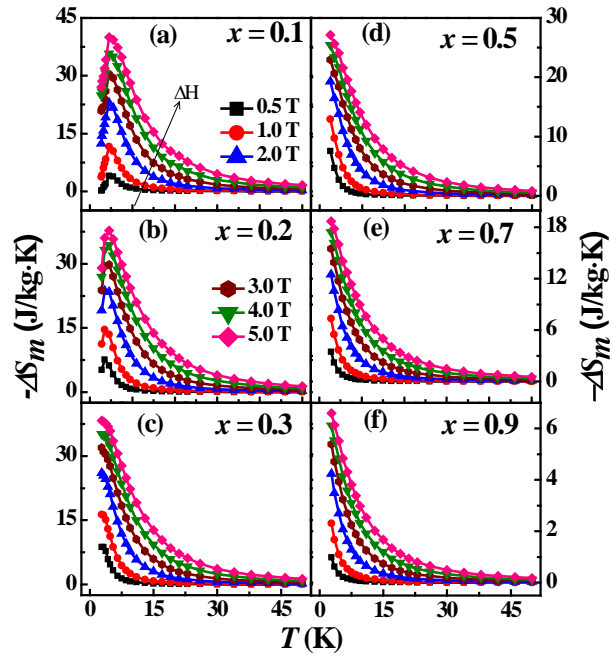

Fig. 5 Km Rubi et al.

**Fig 5**. Temperature dependence of the magnetic entropy change ($-\Delta S_m$) for (a) $x = 0.1$, (b) 0.2, (c) 0.3, (d) 0.5, (e) 0.7 and (f) 0.9 for a field change of $\Delta H$ = 0.5, 1, 2, 3, 4 and 5 Tesla.

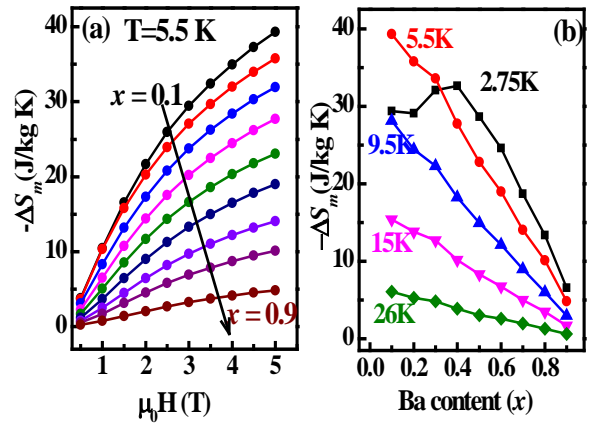

Fig. 6 Km Rubi *et al*.

**Fig 6**. (a) Field dependence of $-\Delta S_m$ at 5.5 K for all the compositions ($x$). (b) Composition ($x$) dependence of $-\Delta S_m$ at $T$ = 2.5 K, 9.5 K, 15 K and 26 K.